\documentstyle[12pt,aasms4]{article}
\begin{document}
\setcounter{totalnumber}{20}
\renewcommand\floatpagefraction{.9}
\renewcommand\topfraction{.9}
\renewcommand\bottomfraction{0}
\title{
THE APPARENTLY NORMAL GALAXY HOSTS FOR TWO LUMINOUS QUASARS
\footnote{Based on observations with the NASA/ESA Hubble Space
Telescope,
obtained at the Space Telescope Science Institute,
which is operated by the Association of Universities for
Research in Astronomy, Inc., under NASA
contract NAS5-26555.\smallskip}}
\author{John N. Bahcall, Sofia Kirhakos}
\affil{Institute for Advanced Study, School of Natural Sciences,
Princeton, NJ~08540}
\centerline{and}
\author{Donald P. Schneider}
\affil{Department of Astronomy and Astrophysics, The Pennsylvania
State University, University Park, PA 16802}

\begin{abstract}
HST images (with WFPC2) of PHL~909\ ($z = 0.171$)
and PG~0052$+$251\ ($z = 0.155$)
show  that these
luminous
radio-quiet quasars each occur in an apparently normal host galaxy.
The host galaxy of PHL~909 is
an elliptical  galaxy  ($\sim$ E4) and the  host of PG~0052$+$251 is a
spiral ($\sim$~Sb).  Both host galaxies are several tenths of a
magnitude brighter
than $L^*$, the characteristic Schechter luminosity of field galaxies.

The images of PHL~909 and PG~0052$+$251, when compared with  HST images of
other objects in our sample of 20 luminous, small-redshift ($z \leq
0.30$) quasars, show that luminous quasars occur in a variety of
environments.
The local environments of the luminous quasars range
from luminous ellipticals, to
apparently normal host galaxies, to complex
systems of interacting components, to faint (and as yet  undetected)
hosts.

The bright HII regions of the host galaxy of PG~0052$+$251 provide an
opportunity to measure directly the metallicity of the host of a
luminous quasar, to establish an upper limit to the mass of the
nuclear AGN (i.e., the putative black hole source),
and to test stringently the cosmological hypothesis
that the galaxy  and the quasar are both at the distance
indicated by the quasar redshift.

The   moderately-luminous  host galaxies of PHL~909 and PG~0052$+$251
are obvious on
the HST images. Normalizing the limits of detectability using
short exposures in which  the host galaxies of PHL~909 and PG~0052$+$251
are easily observed, we estimate that we could have detected similar host
galaxies as
faint as $0.5$ magnitudes less than $L^*$ in the longer-exposure
HST images that have
not yet shown host galaxies.
The details  of the PSF subtraction are unimportant
for the determination of the host galaxy morphologies and
luminosities;
the major and minor axes measured by subtracting very different
stellar PSFs are the same to $\pm 5$\% and the host galaxy magnitudes
are the same to $\pm 0.1$ mag.

\end{abstract}

\section{INTRODUCTION}
\label{introduction}
We present
Hubble Space Telescope (HST)
images of PHL~909 and PG~0052$+$251 that show that these two
quasars reside in apparently normal galaxies; these images provide
unambiguous  evidence
that luminous radio-quiet quasars can be found in prosaic environments.
The evidence is displayed in the accompanying seven figures and the
results are summarized in Table~\ref{characteristics}.  The reader
is urged to at least glance at the figures before proceeding.

PHL~909 and PG~0052$+$251 are two  of the 20
luminous, nearby quasars whose galactic hosts and local environments
are being studied in our HST imaging program.
The  images of the first eight quasars
in this sample have been presented  previously, along with a detailed
discussion of the analysis procedures
(see Bahcall, Kirhakos,
\& Schneider 1994, 1995a; hereafter Paper~I and Paper~II).
In addition, the quasar PKS~2349$-$014 has been caught in the act of what
appears to be a  collision with diffuse galactic material
(Bahcall, Kirhakos,
\& Schneider 1995b, Paper~III). The HST images of 3C~273 provide
remarkably detailed information on the optical counterpart of the radio
jet (Bahcall et al. 1995c, Paper IV).

All of the quasars in the sample have high luminosities and low redshifts;
they were chosen from the   V\'eron-Cetty
\& V\'eron~(1993) catalog and have $z \leq 0.30$ and $M_V
\leq -  22.9$ for $H_0 = 100$ km~s$^{-1}$ ${\rm Mpc}^{-1}$ and $\Omega
_0 = 1.0$. These cosmological parameters are used throughout the present
paper.
The redshift of PHL~909 is $z = 0.171$ and  the apparent
visual magnitude is $V = 15.7$ ($M_V = -22.9$).
The redshift of PG~0052$+$251 is $z = 0.155$ and  the apparent
visual magnitude is $V = 15.4$ ($M_V = -23.0$).
Both PHL~909 and PG~0052$+$251 are radio-quiet (V\'eron-Cetty
\& V\'eron~1993; Kellerman et~al. 1989;
Condon, et~al. 1981).

This paper is organized as follows.  The observations are described in
\S~\ref{obssection}.  The images and the measurements
are presented in \S~\ref{ressection}.
The principal results and conclusions of this paper are summarized and
discussed in \S~\ref{summary}.

There have been a number of previous investigations of the nebulosity
around PG~0052$+$251 using ground-based observations (see, e.g., Boroson, Oke,
\& Green 1982; Boroson, Person, \& Oke 1985; Stockton \& MacKenty
1987; Hutchings, Janson, \& Neff 1989; Dunlop, et~al. 1993;
and McLeod \& Rieke 1994), while relatively few studies
have been made of the host environment of PHL~909\ (see Gehren, et~al.
1984; Dunlop et~al. 1993).  The HST results will be compared with the
previous ground-based results in \S\ref{groundcomparison}.

\section{OBSERVATIONS}
\label{obssection}

The quasars PHL~909 and PG~0052$+$251 were observed on October 17, 1994, and on
December 4, 1994, respectively, for 1400s, 500s, and 200s
with the Wide-Field Camera (WF3) of HST.
All of the exposures were taken with  the
wide-band visual filter, F606W.
The center-of-light of the quasar PHL~909 was placed
$2\farcs6$ from the center of
WF3 and the center-of-light of PG~0052$+$251 was $4\farcs9$
from the center of
WF3. The scale of WF3 is 0\farcs0996\  pixel$^{-1}$.
For further information about the WFPC2,
see Burrows~(1994), Trauger et~al.~(1994), and Holtzman et~al.~(1995a,b).
Further details of the observational
procedures are given in Paper~I and Paper~II.

\section{RESULTS}
\label{ressection}

In this section, we first discuss
in \S~\ref{hostgalaxies} the images of the host galaxies of
PHL~909 and PG~0052$+$251 and then determine their
principal quantitative characteristics.
Next  we present in \S~\ref{fieldgalaxies} the positions and apparent
magnitudes of the galaxies that are projected relatively close to
PHL~909 and PG~0052$+$251.  Finally, we give in \S~\ref{sectionHII}
the measured
apparent magnitudes and projected separations
of the ten brightest HII regions in the host galaxy of PG~0052$+$251.

We report measured F606W magnitudes on the
HST photometric scale established by Holtzman et~al.~(1995b).
For convenience in comparison with standard galaxy magnitudes,
we convert the F606W magnitudes to $V$ magnitudes using the
calculations of k-corrections and the relative sensitivies of different
bands calculated  by   Fukugita, Shimasaku, \& Ichikawa (1995).
To determine the offsets between $V$ and F606W, we interpolate
in redshift  between entries in
their Table~3 and their Table~6.
For an elliptical galaxy at the redshift
of PHL~909, we find $V - m({\rm F606W}) = 0.44$ mag. For an Sb
galaxy at the
redshift of PG~0052$+$251, we find $V - m({\rm F606W}) = 0.32$ mag.
We will use these
conversions below when transforming from F606W to $V$ magnitudes.

None of the major
conclusions of this paper are affected significantly by
corrections to the inferred $V$ magnitudes caused by
k-corrections, magnitude offsets, and photometric zero-point differences.

Cosmic ray subtraction and pipeline STScI flatfielding are the
only processing performed on the
HST images used to construct the seven figures of this paper, except
where explicitly stated otherwise.
Cosmic ray removal was a relatively easy task as the three exposures
of a given quasar were aligned to an accuracy of better than 0.3 pixels.
Cosmic rays were identified by a pixel-by-pixel comparison of pairs of
images; the intensity of a pixel containing a cosmic ray was replaced
by the scaled value of the intensity of the pixel in the other image.

\subsection{Host Galaxies of PHL~909 and PG~0052$+$251}
\label{hostgalaxies}

We first present and analyze the images of the
host galaxy of PHL~909 (in \S~\ref{subhostphl}) and then discuss the host
galaxy of PG~0052$+$251 (in \S~\ref{subhostpg}).  Following
the procedure that is
frequently used
with ground-based observations of quasar hosts, we next discuss (in
\S~\ref{suboned}) the
one-dimensional profiles of both galaxies. Finally, we demonstrate
by using different procedures (in \S~\ref{insensitivity})
that the precise way
we subtract the stellar PSF does not affect significantly the measured
morphology or photometry of the host galaxies.
Finally, we summarize (in \S~\ref{subhostsummary})
the information that is obtained on the two host
galaxies.

\subsubsection{The Host Galaxy of PHL~909}
\label{subhostphl}

Figure~\ref{figphl} displays the environment of PHL~909.
 The  host galaxy of PHL~909
is obvious even in the raw data of the HST image.

The upper panel of
Figure~\ref{figshort} presents the short exposures, 500s and 200s,
of PHL~909. The host galaxy is easily visible on the 500s image and
can also be recognized on the shortest exposure, 200s.

Figure~\ref{phlsub} shows the host
galaxy in somewhat more detail since a best-fitting stellar PSF
(representing the quasar nucleus) was subtracted from the original
image (see Paper~I and Paper~II for a description of the PSF and the
subtraction procedures).
To a limiting  surface brightness of $\mu = 24.5
{}~{\rm arcsec}^{-2}$ (used here and in later discussions of PHL~909),
the major axis of the host  galaxy is  $18''$ ($34$~kpc) and the
the minor axis is $10''$ ($19$~kpc). The position angle of the major
axis is $138^\circ$.
The measured sky brightness in the PHL~909 image was 190 detected photons
per pixel in 1400s, with a standard deviation of 14 photons,
which corresponds to a sky brightness of $22.1$ mag arcsec$^{-2}$.

We have performed aperture photometry (between $0\farcs5$ and $10\farcs0$)
on the PSF-subtracted image of
PHL~909.
We find $m({\rm F606W}) = 17.3 \pm 0.3$ and
$M_V = -20.9$, which is more than one  mag fainter  than the
brightest elliptical galaxies in rich clusters
(cf. Hoessel \& Schneider
1985; Postman \& Lauer 1995).

The host  galaxy of PHL~909
is, at least in its appearance in this F606W image, smooth
and relatively featureless,
similar to a normal E4 galaxy.
Within the inner $1''$, the surface brightness of the host galaxy
appears to be somewhat higher on one side of the quasar compared to
the other side.
In addition, there are some arc-like features
at $\sim 2''$ from the center of
light of the quasar. We are not able to determine from the available
data whether these features are artifacts.
A measurement of the surface brightness along the major axis of
PHL~909
does not show evidence for a break in the light distribution (upper
limit $\sim 0.15$~mag) that might be associated with the end of a disk.
We therefore tentatively adopt the
classification E4.
Additional imaging of PHL~909
with HST using different filters with the WF3 (to determine the host
galaxy colors) and
separate exposures with the PC2 (to determine better the innermost
structure of the host galaxy), as well as comparable CCD images of
classical nearby galaxies,  would be useful in making
a more
definitive classification  of the host galaxy.

\subsubsection{The Host Galaxy of PG~0052$+$251}
\label{subhostpg}

Figure~\ref{figpg}
shows a beautiful spiral galaxy in which the quasar
PG~0052$+$251 apparently resides.
Bright HII regions are visible in the spiral
arms.
The figure displays the  1400s HST image with the cosmic rays
removed.

The lower two panels of Figure~\ref{figshort} demonstrate
the extrordinary power
of the Wide-Field camera  for discovering host galaxies of quasars.
The host galaxy of PG~0052$+$251 is easily visible on all three of the
HST images, including the very short 200s exposure.

Figure~\ref{pgsub} shows  the host galaxy after a best-fit stellar PSF
(representing the quasar nucleus) is subtracted from the original image.
To a limiting  surface brightness of $\mu = 24.8
{}~{\rm mag~arcsec}^{-2}$ (used here and in later discussions of
PG~0052$+$251),
the major axis of the host  galaxy is  $18''$ ($32$~kpc) and the
the minor axis is $12''$ ($21$~kpc). The position angle of the major
axis is $173^\circ$.
The sky brightness in
the 1400s PG~0052$+$251 image is $22.3$ mag arcsec$^{-2}$.

Comparing the shape of the  spiral arms and the rather
prominent dust lanes with figures in the Hubble Atlas, we suggest that
this host galaxy resembles a normal Sb galaxy. The inclination angle
between the plane of the host
galaxy and our line of sight is $ 50 \pm 5$\arcdeg.

We have also performed aperture photometry (between $0\farcs5$
and $10\farcs0$)
on the PSF-subtracted image of
PG~0052$+$251. We find $m({\rm F606W}) = 17.0 \pm 0.3$ and
$M_V = -21.1$, which is
$0.6$ mag brighter  than an $L^*$ galaxy
in the field (Schechter 1976; Kirshner et~al. 1983;
Efstathiou et~al. 1988).

\subsubsection{One-Dimensional Surface Brightness Profiles}
\label{suboned}

Figure~\ref{figprofiles} shows the one-dimensional (azimuthally-averaged)
surface brightness profiles of the host galaxies
of PHL~909 and PG~0052$+$251, as
measured on the HST images after subtraction of a best-fit stellar PSF
(cf. Figure~\ref{phlsub} and Figure~\ref{pgsub}). The filled circles
show the data and the best-fitting de Vaucouleurs profile is
represented by a continuous line.  The best-fitting exponential
profile is indicated by a dotted line.

The best-fitting de Vaucouleurs profile fits very
well the host galaxy of PHL~909 and yields an effective radius of
$2\farcs0$.  The total de Vaucouleurs magnitude of the host galaxy of
PHL~909 is $m$(F606W) $= 16.8$,
somewhat brighter, as expected,
than the measured aperture magnitude of $17.3 \pm 0.3$ mag.
The average surface brightness at the effective radius
is $\mu = 21.8 \pm
0.2$ mag~arcsec$^{-2}$.  The exponential profile is not a good fit to
the light distribution of the host galaxy of PHL~909.

The measured one-dimensional profile of the host galaxy of PG~0052$+$251
is very interesting.
For this spiral host galaxy,
the bottom panel of Figure~\ref{figprofiles} shows that the best-fitting
de Vaucouleurs profile fits well the measured surface brightness
down to a surface brightness level of about $23$ mag~arcs$^{-2}$.
Below $23$ mag~arcs$^{-2}$, the de Vaucouleurs
profile falls more rapidly than the measured profile.
The exponential disk profile
does  not reproduce well the measured
values in the inner  regions, but does fit the measured profile in
the outer regions about as well as the de Vaucouleurs profile.
These results are  consistent with what should
be expected, upon reflection,
for a  spiral galaxy in which the light in the inner
regions is dominated by  a de Vaucouleurs bulge and  the light in
the outer regions comes primarily from  an exponential
disk.

The exponential profile has a scale
length of $1\farcs4$ and a total magnitude of $m{\rm (F606W)} = 16.8$.  The
exponential fit has a total magnitude slightly brighter than the aperture
magnitude of $17.0 \pm 0.3$.

We have estimated an apparent magnitude for the bulge component of the
host galaxy of PG~0052$+$251 in the following manner.
We subtracted the best-fit
exponential surface brightness
profile given in Figure~\ref{figprofiles}, which fits the
observed light distribution beyond $1''$, from the total
surface brightness profile and assigned the residual light to a bulge.
We find in this way a spheroid magnitude of $m{\rm(F606W)} = 19.9 \pm
0.2$. The spheroid is about $3.1$ mag fainter than the disk,
which is consistent with what is known about the  disk and
spheroid of the Galaxy (see, e. g., Bahcall, Schmidt, and Soneira
1983).

The HST images of PG~0052$+$251 show that one must use caution in
interpreting the results of one-dimensional
profile fits to the residual light measured after subtracting a
stellar PSF from a quasar image.  Many authors have, working with data
obtained from ground-based telescopes,  used such studies
to infer the morphology of the host galaxies of quasars.  However,
for the spiral host of PG~0052$+$251
the bottom panel of  Figure~\ref{figprofiles} shows that
a de Vaucouleurs profile fits the measured light
distribution  as well as an exponential profile.

The parameters derived from the fits to the radial profiles are not
sensitive to how the points are weighted.  We fit the profiles of PHL~909
and PG~0052+251 using four weighting schemes: uniform, $({\rm radius})^{1/2}$,
$\sigma^{-1}$, and $\sigma^{-2}$, where $\sigma$ is the uncertainty in the
in the measurements of the individual points.  The four derived luminosities
for each galaxy varied by only 0.1~mag.  The scale length for the PG~0052+251
host changed by less than 0.1" using the various weightings; the scale length
for PHL~909 ranged from 1.4\arcsec to 2.0\arcsec.

\subsubsection{Insensitivity to Stellar PSF Subtraction}
\label{insensitivity}

How sensitive are the measurements of the morphology and the
photometry of the
host galaxies to the details of the PSF subtraction of the stellar
(nuclear) quasar?  In order to answer this question, we have carried
out the measurements in  different ways and compared the results.
Altogether, we measured the characteristics of the host galaxies in
four different ways: 1) we subtracted our best stellar PSF from the
total image, minimizing the residuals in the  diffraction spikes
 (designated as Best PSF in Table~\ref{dependence});
2) we subtracted a much-inferior stellar PSF and then repeated the
measurements (designated as Bad PSF);
and 3) we measured the major and minor axes of the host galaxies without
subtracting any stellar PSF (designated as No Subtraction).

The results of these measurements are summarized in
Table~\ref{dependence} and described below.

The values obtained by subtracting our best PSF have been presented in
the previous discussion of \S~\ref{hostgalaxies}; the values obtained are
given in Table~\ref{dependence} in the first and fourth rows.

We constructed an alternate PSF from a
star image that was taken about five months before, and in a
different CCD, than
the two quasar images.
Moreover, the exposure level of this image was insufficient to accurately
determine the outer regions of the PSF.  We will refer to this PSF as the
``Bad PSF," following the nomenclature of Table~\ref{dependence}.

Subtracting the Bad stellar PSF from the image of PHL~909, we
found an effective radius of
$2\farcs4$ ($0\farcs4$ more than with our best PSF) and a total
magnitude of $m{\rm (F606W)} = 16.6$  ($0.2$ mag brighter than obtained with
our best PSF subtraction).  The major and minor axes measured with the
Bad PSF subtraction were $19''$ and $10''$, respectively, compared
to $18''$ and $10''$ with our best PSF. For PG~0052$+$251, we found with the
 Bad
stellar PSF, an exponential scale length of $1\farcs5$ ($0\farcs1$
more than
with our best PSF) and a total magnitude of $m{\rm (F606W)} = 16.8$ (in
agreement with our best PSF). The major and minor axes measured with the
 Bad PSF subtraction were $18''$ and $11''$, respectively, compared
to $18''$ and $12''$ with our best PSF. These results are given in
rows two and five of Table~\ref{dependence}.

For PHL~909, the major axis without PSF subtraction was measured to be
 $17''$
compared to $18''$ with our best stellar PSF subtraction; the minor
axis was
measured to be $10''$ without PSF subtraction, which is to be compared
to $10''$
with PSF subtraction.  For PG~0052$+$251, the major and minor axes measured
without PSF subtraction were $18''$ and $11''$, respectively, compared
to $18''$ and $12''$ with out best PSF subtraction.
These results are presented in rows three and six of
Table~\ref{dependence}.

We conclude that, within the class  of procedures we have
considered,  the details of the subtraction process are
not important for measuring the luminosities or the morphological
characteristics of the host galaxies.

\subsubsection{Summary of Host Galaxy Characteristics}
\label{subhostsummary}

Table~\ref{characteristics} summarizes the principal characteristics
of the host  galaxies of PHL~909 and PG~0052$+$251, as determined from
the HST observations.  The centers of the
galaxies were determined by locating the pixels of given surface
brightnesses, fitting ellipses to this distribution, and identifying
the center of the galaxy with the center of the ellipse.  The uncertainties
in the positions of the centers of the galaxies were estimated by the
dependence of the ellipse centers with surface brightness.
The center of the host galaxy and the center of the quasar nuclear
light are  coincident for both PHL~909 and PG~0052$+$251 to within the
accuracy with  which we can locate the galactic centers.
The measurement error for the galactic centers   is 0\farcs3
for PG~0052$+$251 and 0\farcs5  for PHL~909.

\subsection{Galaxies in the Fields of PHL~909 and PG~0052$+$251}
\label{fieldgalaxies}

No very close
galactic companions
 (projected separation less than 3\arcsec, cf. Paper~II)
to PHL~909 are visible in the HST images of
this quasar.
There are two  faint stellar images projected  close to  PG~0052$+$251, the
closest is $4\farcs2$ north-west of the quasar nucleus and the
more distant is $6\farcs5$  south of the quasar nucleus.  Their
apparent magnitudes are, respectively, $23.6$ and $21.3$.

Table~\ref{phltable} and Table~\ref{pgtable} list the
aperture magnitudes (apertures of $0\farcs5$ to $4\farcs0$ as appropriate)
of galaxies in the
fields of PHL~909 and of PG~0052$+$251 that are brighter than
$m{\rm (F606W)} = 22.5$ and
that have a projected separation
of less than 70~kpc (at the quasar redshift),
from the center-of-light of the quasar.
The F606W magnitudes  given in Table~\ref{pgtable} are typically $0.7$
mag brighter than the $g$-magnitudes given in Kirhakos et al. (1994);
within the measurement uncertainties ($\sim 0.3$~mag),
this is consistent with what would be expected from the filter offsets
and k-corrections (cf. Fukugita et al. 1995).

It would be useful to obtain redshifts for
these galaxies.  When the galaxy
redshifts are available, one can investigate
the extent of the galactic halos with HST ultraviolet absorption-line
spectra of the quasars.

It is possible  that one or more of the galaxies listed in
Table~\ref{phltable} or Table~\ref{pgtable} is associated with either
PHL~909 or PG~0052$+$251.  With our definition of aperture magnitudes,
the average galaxy density brighter than 22.5 mag in the
four CCDs in which the quasars do not appear is $7 \times 10^3$
galaxies per square degree.  For consistency,
we consider galaxies that are within the same separation, 25~kpc,
from the
quasar centers of light as was adopted in the discussion of companion
galaxies  in Paper~II.
With this observed  average  density of galaxies, there is
a 16\% chance that by accident there would be, as observed, two
galaxies that are projected within 25~kpc of the quasar centers of
light\footnote{
After this paper was accepted for publication, we were informed by T.
Boroson of an important observing project by
T. Boroson, J. Dunlop, and D.
Hughes in which they have
obtained redshifts for two of the galaxies in the PHL~909 field
(see Table~\ref{phltable}).  For galaxy C, they
find (Boroson 1995) $z =
0.102$ and for galaxy F they find $z = 0.169$.}.  (This estimate of
the probability may be somewhat of an overestimate since the
average galaxy
density in the four CCDs in which quasars do not appear may be
enhanced over the field density.)

\subsection{The Brightest HII Regions in the Host Galaxy of PG~0052$+$251}
\label{sectionHII}

Figure~\ref{pgHII} shows the eleven brightest HII regions that we have
found in the host galaxy of PG~0052$+$251.  The measured apparent magnitudes of
these HII regions range from $m{\rm (F606W)} = 22.9$ to
$m{\rm (F606W)} = 24.9$.
The  brightest HII region  is marked as  ``j'' in
Figure~\ref{pgHII} and is discussed in \S~\ref{discussHII}.

Table~\ref{HIItab} presents the measured characteristics of the HII
regions shown in Figure~\ref{pgHII}, including the apparent magnitudes,
the distance, $d$, in
arcseconds of the HII region from the quasar nucleus,
and the offsets in right ascension and declination between each HII
region and the quasar nucleus.  The aperture magnitudes given in
Table~\ref{HIItab} were measured with apertures of 0\farcs3--$1\arcsec$;
the typical magnitude uncertainties are
$\pm 0.3$~mag but are somewhat larger for those regions with  $d < 1\farcs4$.

\section{DISCUSSION}
\label{summary}

In this section, we summarize in \S~\ref{hostconclusions}
 the conclusions about host galaxies of luminous quasars
 that follow from the analyses of HST images studied in this paper and in
previous papers in this series.
We describe in \S~\ref{validation} how the detection of the host
galaxies of PHL~909 and PG~0052$+$251 validates the techniques we have
used in analyzing HST observations.
Next we compare  in
\S~\ref{groundcomparison} the HST results with previous ground-based
studies of PHL~909 and PG~0052$+$251.  Then, we discuss in \S~\ref{discussHII}
the HII regions that are found in the spiral arms of PG~0052$+$251 and stress
the importance of studying these regions spectroscopically.
Finally, we point out in \S~\ref{rotation} that it may be feasible to
obtain a direct upper limit on the mass of the nuclear region of PG~0052$+$251
by measuring a rotation curve for the host galaxy.

\subsection{What Can We Conclude About the Host Galaxies of Luminous
Quasars?}
\label{hostconclusions}

The principal conclusion from the results presented in this and in the
previous papers in this series (Papers~I--IV) is that luminous quasars
exist in a variety of environments.  There are quasars for  which
no definitive evidence for host galaxies is seen to the limit of the
sensitivity obtained so far, which is
typically of order $L^*$ ($M_V(L^*) = -20.5$), or a magnitude
fainter, where $L^*$ is
the characteristic Schechter magnitude of field galaxies (Kirshner
et~al.
1983; Efstathiou, Ellis, \& Peterson 1988). Examples of quasars in
this category include (see Paper~I and Paper~II) PG~0953$+$414,
PG~1202$+$281, PKS~1302$-$102, and 3C~323.1.  There are also two
quasars in which host elliptical  galaxies are clearly detected,
namely, 3C~273 (Paper~II) and PHL~909.  Finally, PG~0052$+$251 exists  in a
spiral galaxy and PKS~2349$-$014 is embedded in  a
complex environment, including a large, off-center nebulosity
and thin (possibly tidal) wisps (see Paper~III).

In the sample of luminous quasars that we have studied,
radio-quiet and radio-loud quasars are not uniquely
distinguished by their host
galaxies.  There are both radio-loud (e.g.,
PKS~1302$-$102, and 3C~323.1)
and radio-quiet (e.g., PG~0953$+$414 and PG~1202$+$281)  quasars for which we
have not detected the host galaxies and there is both a radio-quiet (PHL~909)
and a
radio-loud (3C~273) quasar for which we have detected
a host elliptical  galaxy.\footnote{
It is an interesting and instructive
historical exercise to try to trace the early
development of the consensus view, based upon ground-based
observations, that radio-quiet
quasars are  in spiral galaxies and radio-loud quasars are  in
elliptical galaxies.
The motivating  analogy may have been
that Seyfert galaxies are predominantly
early-type spirals and strong radio galaxies are predominantly
ellipticals.
Landmark studies
include Sandage (1972), Balick and Heckman (1982),
Malkan, Margon, and Chanan (1984), and Malkan (1984).}

Do all luminous quasars reside in luminous galaxies?  No,
the results presented here for PHL~909 and PG~0052$+$251 provide further
evidence supporting the  conclusion stated in Paper~I and Paper~II that
some luminous nearby quasars do not reside in particuarly
luminous  galaxies.
We easily see moderately-bright host galaxies for PHL~909 and
PG~0052$+$251 with
the same exposures that do not reveal any host galaxies for other
quasars with similar nuclear properties.
For example, PG~1202$+$281\  (studied in Paper~I and Paper~II)
has a redshift $z = 0.165$, similar to (and
intermediate between) the redshift
of PHL~909\ ($z = 0.171$) and the redshift of PG~0052$+$251\ ($z = 0.155$).
Also, the
apparent magnitude of the quasar nucleus is similar for PG~1202$+$281\ ($V
= 15.6$) and for PHL~909\ ($V =15.7$) and PG~0052$+$251\ ($V = 15.4$).
The host galaxies ($M_V \sim -21$) for PHL~909 and for PG~0052$+$251 are
visible on the HST
images with no special data processing, whereas we
could not detect any convincing evidence for a host galaxy on similar
HST images of PG~1202$+$281. The upper limit on the brightness of a model
spiral host galaxy (similar in morphology to that seen in PG~0052+251)
was set at $M_V = -19.6$ in Paper~II.

Of course, the quasars for which we have not yet detected host
galaxies may reside in  fainter galaxies.  We showed in
Paper~II, for example, that our results on the first eight quasars we
studied were consistent with the host galaxies being described by a
Schechter luminosity function, provided that the cutoff at faint
luminosities was at least as faint as $0.5 L^*$.

The ultraviolet ``big bump'' that appears in   many quasar spectra
has been suggested to be a signature of thermal radiation from the
accretion disk around a central black hole (Malkan 1983).  It is of
interest that the two elliptical host galaxies that we have detected
correspond to very different types of ``big bumps.'' PHL~909
has the weakest bump
in the sample of McDowell et~al. (1989) and 3C~273 has
one of the strongest ultraviolet bumps.

\subsection{Validation of the Analysis Techniques: PHL~909 and
PG~0052$+$251}
\label{validation}

There is an old adage: ``The proof of the pudding is  the eating.''
The validation of the techniques that we have used is  the detection of
the host galaxies of PHL~909 and PG~0052$+$251PH
and  the measurement
of their properties.
For faint very extended nebulosity, the validation of the technique is in
the detection of the emission around PKS~2349$-$014 (see Paper~III).

The short, $200$s, exposures provide  an empirical calibration for the
detectability of host spiral or elliptical galaxies in the seven times
longer
exposures ($1400$s) of eight quasars studied with similar
techniques (see
Paper~II).  The relative increase in sensitivity is bounded
by two extremes:
1)~read-noise limit (2.1~mag increase in sensitivity), and
2)~sky-noise
limit
(1.05~mag increase in sensitivity).   The sky-noise approximately
equals the
readout-noise
in the $200$s exposures in this filter, so
the limiting sensitivity for host galaxies in $1400$s is about
1.4~mag fainter than the
absolute magnitudes measured in this paper for the
host galaxies of PHL~909 and PG~0052$+$2251.  These estimates
must be  modified
somewhat by
the fact that the host galaxies of PHL~909 and   PG~0052$+$251  are well
above the detection thresholds on the $200$s
exposure, which is perhaps compensated for by the fact that
we have neglected additional saturated pixels and scattering that are
produced
by the quasar nucleus in the longer exposures.  We conclude that host
 galaxies
like PHL~909  and PG~0052$+$251 could be detected to a limiting
magnitude of about
$M_V$~=~$-20.0 \pm$~0.5 on $1400$s exposures.  This value is in
good agreement with the estimates of our limiting sensitivity
determined from simulations in Paper~II.

We have shown in \S\ref{insensitivity} and in Table~\ref{dependence} that
the inferred
characteristics of the host galaxies of PHL~909 and PG~0052$+$251 are
insensitive
to the details of the different
procedures that we use to subtract  a stellar PSF
from the HST images.  The difference in the host galaxy magnitudes
 derived using our Best PSF and a Bad PSF is $0.2$~mag for PHL~909 and
$0.0$~mag for PG~0052$+$251. The measured major and minor axes are
the same, to
10\% or better, for PHL~909 and PG~0052$+$251, independent of whether the
measurements are made with the Best PSF, the Bad PSF, or no
subtraction at all.
These tests reinforce
the result  obtained  in Paper~II, where we
showed that we obtained the same apparent magnitudes
(to an accuracy of $\pm 0.17$ mag) for model host galaxies that were
fit to the observations in two very
different ways. In Paper~II, we minimized the residuals in the fits by
either using all of the emission in annuli centered on the quasar
images (the common practice in analyzing ground-based observations)
or by first matching as well as possible the diffraction spikes
in the quasar and the stellar images.

\subsection{Comparison with Ground-based Studies}
\label{groundcomparison}

The lower panel of Figure~\ref{figprofiles} shows that a de
Vaucouleurs profile provides an excellent fit to the
azimuthally-averaged  light
distribution of the host spiral galaxy of PG~0052$+$251 over a large range in
angular distance, from $0\farcs3$ to about $5\farcs0$.  Over this angular
range, the observed surface brightness varies by more than
$4.5$ mag. In fact, the de
Vaucouleurs fit to the azimuthally-averaged light distribution
is a better fit than the exponential disk model over nearly all the
measured range (cf. Figure~\ref{figprofiles}).  The full
two-dimensional light distribution is required to see that the host of
PG~0052$+$251 is a spiral, not an elliptical, galaxy.

Since it has been standard practice to infer the morphological type of
host galaxies of quasars
from azimuthally-averaged ground-based observations,
the results shown in the lower panel of Figure~\ref{figprofiles}
suggest that the inferences regarding the morphological types of host
galaxies  based upon
azimuthally-averaged profiles should be reexamined.

The host galaxy of PG~0052$+$251 has been the subject of several previous
ground-based studies.  The first such study for PG~0052$+$251 of which we are
aware is a spectroscopic investigation by Boroson, et~al.
(1982), who concluded from measurements of the nebulosity surrounding
the stellar source that the host was probably a spiral galaxy with
(for our choice of $H_0$) $M_V = -20.5$; this result is in satisfactory
agreement with our HST aperture magnitude of $M_V = -21.1$.
{}From near infrared ground-based observations,
McLeod \& Rieke (1994) estimated an  $H$-band magnitude for
the host galaxy: $M_H = -24.0$.
This result is in good agreement with our measured absolute magnitude
at $V$, if we adopt a typical value of $V - H$ that applies for
ordinary galaxies ($V - H \sim 3$, cf. McLeod \& Rieke 1994).

There has been relatively little ground-based imaging of PHL~909.
Gehren, et~al. (1984) estimated an $r$-band
absolute magnitude for the host galaxy of $M_r = -21.5$,  which is
consistent
with our measurement of $M_V = -21.4$.  In addition, Gehren et~al. suggest
the presence of diffuse emission indicating a tidal
interaction with galaxy B of Figure~\ref{figphl}.
Dunlop et~al. (1993) performed $K$-band observations and suggest a
bright host galaxy, $M_K = -24.2$, which is consistent with our
measurement if $V - K \gtrsim +3$~mag.  With regard to morphology,
Dunlop et al. propose that the host of PHL~909 may extend
towards Galaxy B.
The HST images show (see Figure~\ref{figphl}) that the quasar and
Galaxy B are not connected by diffuse emission that is detectable in
our relatively-deep exposures (limiting surface brightness of about 25
mag arcsec$^{-2}$).  It is possible that the proximity of the two bright
sources, PHL~909 and Galaxy B, on the ground-based images may---because of
limited angular resolution---have given rise to a
misleading impression of tidally-connected emission.

As a further
indication of the kind of difficulty that may face ground-based imaging
of some quasar hosts,
we note that
Hutchings, et~al. (1989) also detected (in deep $B$ and $R$ images)
the  bright   region ``j'' (see Figure~\ref{pgHII}),
but suggested that it was a
secondary nucleus of the quasar.
They also  suggested that
the bright stellar object about $6\farcs5$ south-east of the quasar (see
Figure~\ref{pgsub}) was a third quasar nucleus.
Dunlop et~al. (1993), observing in
the $K$-band, detected the stellar image and concluded, in agreement
with the  Hutchings et~al. suggestion, that it was a secondary nucleus
linked to the quasar.
The HST data confirm the existence of emission at the locations found
by Hutchings et al. and Dunlop et al.
Given the  location of object ''j'' within a  spiral arm of PG~0052$+$251
and upon the   morphology  of the host galaxy as seen  in the HST
images (cf. Figure~\ref{pgsub}), we conclude that the object `j'' is
almost certainly an HII region and not a secondary quasar nucleus.
The bright stellar object $6\farcs5$ south-east of the quasar is almost
certainly a star and not another quasar nucleus.
On the  HST images, this object
has the light profile of a star and is well-separated
from the quasar nucleus.

\subsection{The HII Regions of PG~0052$+$251}
\label{discussHII}

The discovery  of bright HII regions in the spiral arms of the host
galaxy of
PG~0052$+$251 makes possible   important new observations.
For example,  the region marked  ``j'' in
Figure~\ref{pgHII} has  $m{\rm (F606W)}
\approx 22.9$ and is  at a projected distance of about  $4\farcs1$
from the quasar nucleus; it should be possible to obtain a good
spectrum of this HII region with a large ground-based telescope.
Stockton \& MacKenty (1987) have already shown by narrow-band imaging
in the [OIII] $\lambda 5007$ line that the HII region ``j'' has a
luminous flux in the $30$ \AA\ bandpass centered on redshifted
$5007$~\AA.

It would be of great interest to obtain spectra for as many of
the HII regions as possible
and to measure their element abundances.  This information might
provide significant clues to the history and nature of the system in
which the luminous AGN PG~0052$+$251 resides.  The relative velocities of the
HII regions could also provide valuable information about the dynamics
of the host system.

The HII regions also make possible
direct
tests of the cosmological hypothesis that the quasar and the host
galaxy are both at the distance indicated by the quasar redshift.
According to this hypothesis, the emission lines in the HII regions
should be redshifted by $z{\rm (HII)}= 0.155$.

In principle,
the
apparent magnitudes of the brightest HII regions in the host
galaxy of PG~0052$+$251 can also be used to test whether
the galaxy is at the same redshift as the quasar.
The broad-band measurements of HII regions by
Wray \& de Vaucouleurs (1980)
are potentially useful for this purpose.
In practice, using magnitudes of
the HII regions in just one color to say something about  the  distance
to PG~0052$+$251\ involves  large uncertainties because we
do not know the $B - V$ color of the host galaxy, the intrinsic
reddening of the host galaxy, the $B - V$ colors of the HII regions,
and the total absolute magnitude of host galaxy.
Moreover, the HST filter is different from the traditional $V$ filter
used by Wray \& de Vaucouleurs and the emission lines of the HII
regions in the host galaxy of PG~0052$+$251 are
redshifted to longer wavelengths
than they are  in the calibration made using nearby galaxies.
We also do not know whether the distribution of HII absolute
magnitudes is affected by the presence of the quasar nucleus.
If we nevertheless make a crude,
preliminary estimate based upon the data of Wray \& de Vaucouleurs and
the data presented in the present paper, ignoring everything we do
not know, we
find that the HII regions in PG~0052$+$251 are about  a magnitude brighter
than would be expected on the basis of this oversimplified calculation.

\subsection{A Rotation Curve for PG~0052$+$251?}
\label{rotation}

Finally, we note that
it is  possible to obtain a rotation curve for the host
galaxy of PG~0052$+$251.
The host galaxy is bright enough, $V \sim 17.0$, that a good spectrum
could be obtained with HST and perhaps even with a ground-based
telescope in excellent seeing.
One would expect on the basis of general phenomelogical arguments
(see, e.g., Malkan 1983, Rees 1984, Wandel 1991) that a black
hole in the center of PG~0052$+$251 would have a mass of order $10^{9} M_\odot$
or less, which is  less than the expected mass of
the stars and gas in the inner several kpc of the
host spiral galaxy.  Nevertheless,
the rotation curve would set a direct
upper limit on the mass of the quasar nucleus, a quantity of
fundamental interest.

\acknowledgments
We are grateful
to   T. Boroson, S. Casertano, S. G. Djorgovski M. Fukugita,
P. Hodge, J. Hutchings, B. Jannuzi,
R. Kennicutt, J. MacKenty, M. Malkan, K. McLeod,
J. Miller, J. Ostriker,  and M. Strauss
for valuable discussions, comments, and suggestions.
We are indebted to the referee, T. Boroson, for generously allowing us
to cite the redshifts obtained by Boroson, Dunlop, and Hughes
(footnote 2) and to J. Kormendy,
O. Lahav and to A. Sandage for valuable comments on the
classification of the host galaxies of PHL~909 and PG~0052$+$251.
D. Saxe provided
expert help with
the figures. We would like to thank Digital Equipment Corporation for
providing the DEC4000 AXP Model 610 system used for the
computationally intensive parts of this project.
This work was supported in part by NASA contract
NAG5-1618, NASA grant number NAGW-4452 and grant
number GO-5343 from the Space Telescope Science
Institute, which is operated by the Association of Universities for
Research in
Astronomy, Incorporated, under NASA contract NAS5-26555.
\vfil\eject

\newpage

\begin{figure}[h]
\caption[]{PHL~909 and its environment.  The candidate companion
galaxies within $70$~kpc from the center-of-light of the quasar that
are
brighter than $m{\rm (F606W)} = 22.5$ are labeled ``A--J.''
This image and and all the other figures, except
Figure~\ref{figshort},
shown in this paper were obtained
with a $1400$s exposure using the HST WF3 and the F606W filter.
 \label{figphl}}
\end{figure}

\begin{figure}[h]
\caption[]{The short WF3 exposures of PHL~909 and PG~0052$+$251.
The upper panels show 500s
and 200s images with F606W and WF3 of PHL~909; the lower panels show
500s and 200s exposures of PG~0052$+$251.  The spiral host
galaxy of PG~0052$+$251 is
easily seen on both the 500s and 200s exposures.  The early-type host
of PHL~909 is clearly seen on the 500s exposure and, after-the-fact, can
be recognized on the 200s exposure. The only image processing
performed on these short images was cosmic ray subtraction and
pipeline STScI flatfielding.\label{figshort} }
\end{figure}

\begin{figure}[h]
\caption[]{The host galaxy of PHL~909.  This figure shows the host galaxy
of the quasar PHL~909 after a best-fit stellar PSF was subtracted from
the original image that is shown in Figure~\ref{figphl}.  For this
figure, and for Figure~\ref{pgsub}, the values of the saturated pixels
in the very center of the quasar image were replaced by the average
value of the neighboring pixels for cosmetic purposes.\label{phlsub} }
\end{figure}

\begin{figure}[h]
\caption[]{PG~0052$+$251 and its galaxy environment.  The candidate companion
galaxies within $70$~kpc from the center-of-light of the quasar that
are
brighter than $m{\rm (F606W)} = 22.5$ are labeled ``A--G.''
Two foreground stars are identified by ``s''.\label{figpg} }
\end{figure}

\begin{figure}[h]
\caption[]{The host galaxy of PG~0052$+$251.  This figure shows the host galaxy
of the quasar PG~0052$+$251 after a best-fit stellar PSF was subtracted from
the original image that is shown in Figure~\ref{figpg}.\label{pgsub} }
\end{figure}

\begin{figure}[h]
\caption[]{Azimuthally-averaged profiles of PHL~909 and PG~0052$+$251.
Surface brightness profiles are plotted for the host galaxies versus
the logarithm of the radius in arcsec.  The ordinate is the surface
brightness measured in F606W mag~ arcsec$^{-2}$. The filled circles
represent the data and the continuous line show  the best-fitting de
Vaucouleurs profile.  The exponential disk that best fit the data
outside of 1\arcsec\  is
represented by a dotted line.\label{figprofiles} }
\end{figure}

\begin{figure}[h]
\caption[]{The HII regions  of PG~0052$+$251.
The prominent HII regions are labeled
``a--k.''\label{pgHII} }
\vglue6in
\end{figure}

\vfill\eject

\begin{deluxetable}{lcccccccc}
\tablewidth{0pt}
\tablecaption{Characteristics of Host Galaxies of
PHL~909 and PG~0052$+$251\tablenotemark{a}\label{characteristics}}
\tablehead{
\colhead{Quasar}&\colhead{$z$}&\colhead{$V$}&\colhead{$M_V$}&\colhead{F606W}
&\colhead{$M_V$}
&\colhead{kpc/1\arcsec}&\colhead{Size\tablenotemark{b}}&\colhead{Hubble}\\
&&\colhead{QSO}&\colhead{QSO}&&&&&\colhead{Type}
}
\startdata
PHL~909&0.171&15.7&$-$22.9&17.3&$-20.9$&1.88&1\farcs4&E4\nl
&&&&(16.8)&($-$21.4)\nl
PG~0052$+$251&0.155&15.4&$-23.0$&17.0&$-21.1$&1.75&1\farcs4&Sb\nl
&&&&(16.8)&($-$21.3)\nl
\enddata
\tablenotetext{a}{The upper row for each quasar gives aperture
magnitudes between 0\farcs5 and 10\farcs0.
The lower row gives (in parentheses) the magnitudes computed using a
best-fit de Vaucouleurs profile for PHL~909\ or an exponential profile
for PG~0052$+$251.}
\tablenotetext{b}{Effective radius for PHL~909\ and exponential scale length
for PG~0052$+$251.}
\end{deluxetable}

\begin{deluxetable}{llcccc}
\tablewidth{0pt}
\tablecaption{Dependence of Host Galaxy Parameters
on PSF Subtraction\label{dependence}}
\tablehead{
\colhead{Quasar}&\colhead{Procedure}&\colhead{Total
Mag}&\colhead{Major Axis}
&\colhead{Minor Axis}&\colhead{Size\tablenotemark{a}}\\
&&\colhead{(F606W)}&\colhead{(\arcsec)}&\colhead{(\arcsec)}&\colhead{(\arcsec)}
}
\startdata
PHL~909&Best PSF&16.8&18&10&2.0\nl
PHL~909&Bad PSF&16.6&19&10&2.4\nl
PHL~909&No Subtraction&---&17&10&---\nl
PG~0052$+$251&Best PSF&16.8&18&12&1.4\nl
PG~0052$+$251&Bad PSF&16.8&18&11&1.5\nl
PG~0052$+$251&No Subtraction&---&18&12&---\nl
\enddata
\tablenotetext{a}{Effective radius for PHL~909 and
exponential scale length for PG~0052$+$251.}
\end{deluxetable}

\begin{deluxetable}{ccccr@{\hspace{18pt}}r}
\tablewidth{0pt}
\tablecaption{Galaxies in PHL~909 Field\label{phltable}}
\tablehead{
\colhead{Galaxy}&\colhead{$m$(F606W)}&\colhead{$d$}&\colhead{$d$}
&\colhead{$\Delta\alpha$}&\colhead{$\Delta\delta$}\\
&&\colhead{($\arcsec$)}&\colhead{(kpc)}&\colhead{($\arcsec$)}
&\colhead{($\arcsec$)}
}
\startdata
A&21.4&12.5&23.5&1.4&$-$12.4\nl
B&20.5&15.9&30.0&$-$15.5&3.7\nl
C&19.6&17.8&33.5&8.0&15.9\nl
D&19.6&19.6&36.8&12.2&$-$15.3\nl
E&20.4&20.0&37.5&$-$7.0&$-$18.7\nl
F&18.5&28.2&53.1&$-$1.5&$-$28.2\nl
G&22.2&28.4&53.4&$-$28.4&1.2\nl
H&18.5&30.6&57.5&30.4&$-$3.2\nl
I&21.9&31.7&59.7&$-$31.6&3.0\nl
J&22.2&36.4&68.5&$-$30.9&20.1\nl
\enddata
\end{deluxetable}

\begin{deluxetable}{ccccr@{\hspace{18pt}}r}
\tablewidth{0pt}
\tablecaption{Galaxies in PG 0052+251 Field\label{pgtable}}
\tablehead{
\colhead{Galaxy}&\colhead{$m$(F606W)}&\colhead{$d$}&\colhead{$d$}
&\colhead{$\Delta\alpha$}
&\colhead{$\Delta\delta$}\\
&&\colhead{($\arcsec$)}&\colhead{(kpc)}&\colhead{($\arcsec$)}
&\colhead{($\arcsec$)}
}
\startdata
A&18.8&14.0&24.6&3.5&\llap{$-$}13.6\nl
B&19.5&18.5&32.4&1.8&18.4\nl
C&20.9&22.9&40.0&\llap{$-$}5.9&22.1\nl
D&18.2&27.1&47.3&26.4&5.9\nl
E&21.2&27.3&47.8&\llap{$-$}6.7&\llap{$-$}26.5\nl
F&21.9&32.2&56.3&\llap{$-$}13.3&\llap{$-$}29.3\nl
G&\llap{$<$}20.2&39.1&68.5&\llap{$-$}33.4&\llap{$-$}20.4\nl
\enddata
\end{deluxetable}

\begin{deluxetable}{ccccc}
\tablewidth{11cm}
\tablecaption{HII Regions in the Host Galaxy of
PG~0052$+$251\ \label{HIItab}}
\tablehead{
\colhead{Region}&\colhead{$m$(F606W)}&\colhead{$d$}
&\colhead{$\Delta\alpha$}
&\colhead{$\Delta\delta$}\\
&&\colhead{($\arcsec$)}&\colhead{($\arcsec$)}&\colhead{($\arcsec$)}
}
\startdata
a&23.7&1.3&0.9&0.9\nl
b&23.5&1.3&1.3&\llap{$-$}0.3\nl
c&23.8&1.3&\llap{$-$}0.9&\llap{$-$}0.9\nl
d&23.3&1.4&\llap{$-$}1.3&0.5\nl
e&23.9&1.7&\llap{$-$}0.6&\llap{$-$}1.6\nl
f&24.2&3.1&3.1&0.4\nl
g&24.7&3.3&3.2&\llap{$-$}0.7\nl
h&24.5&3.8&2.2&\llap{$-$}3.1\nl
i&24.9&3.9&1.6&3.6\nl
j&22.9&4.1&3.0&\llap{$-$}2.8\nl
k&24.6&4.3&1.9&3.8\nl
\enddata
\end{deluxetable}
\end{document}